# Electrochemical Lensing for High Resolution Nanostructure Synthesis

*Auwais Ahmed, Peter A. Kottke, Andrei G. Fedorov\**

George W. Woodruff School of Mechanical Engineering, Georgia Institute of Technology, 771 Ferst Dr NW, Atlanta, GA 30332, USA

ABSTRACT. The advancement of liquid phase electron beam induced deposition has enabled an effective direct-write approach for functional nanostructure synthesis with the possibility of three-dimensional control of morphology. For formation of a metallic solid phase, the process employs ambient temperature, beam-guided, electrochemical reduction of precursor cations resulting in rapid formation of structures, but with challenges for retention of resolution achievable via slower electron beam approaches. The possibility of spatial control of redox pathways via the use of water-ammonia solvents has opened new avenues for improved nanostructure resolution without sacrificing the growth rate. We find that ammonia concentration locally modulates reaction kinetics, altering the balance between reducing and oxidizing species, leading to distinct deposition outcomes. The key effect is an 'electrochemical lensing', achieved at an optimum ammonia concentration, in which a tightly confined and highly reducing environment is created locally to enable high resolution, rapid beam-directed nanostructure growth. We demonstrate this unique approach to high resolution synthesis through a combination of analysis and experiment.

INTRODUCTION. Techniques using focused electron beams generally have the potential to lead to the highest resolution nanostructure synthesis. Extensive research has been conducted on the application of the focused electron beam for the fabrication of nanostructures, demonstrating its effectiveness, precision and versatility across nano- and microscales.[1] The utility of Focused Electron Beam Induced Deposition (FEBID) has been shown in material science applications, including nanoelectronics, optics, nanoprobes, plasmonics, and magnetics.[2,3] Beam-based fabrication provides the ability to create nanostructures atom-by-atom on suspended atomically thin substrates with minimal parasitic co-deposition effects.[1] However, this potential for "spot-on" synthesis is yet to be fully realized especially in an environment that meets requirements relevant to practical applications. The most commonly used FEBID approach, gas phase electron beam induced deposition, suffers from low growth rate and impurities due to parasitic carbon entrapment.[3] Increasing growth rates often leads to unwanted deposition, compromising resolution,[4] as there is a general tradeoff between growth rate and resolution. In this work, we describe an approach to resolve this fundamental tradeoff with a technique we call 'electrochemical lensing'. Electrochemical lensing exploits principles of electrochemistry and transport to facilitate high growth rates without sacrificing resolution.

Use of liquids as a precursor for FEBID (LP-FEBID) has been shown to offer higher growth rates, higher purity, and a greater selection of precursors as compared to gas phase precursors.[5] For high resolution FEBID, the process takes place inside a vacuum chamber where rapid evaporation makes the use of liquid phase precursors challenging. Enclosed cells, with electron transparent membranes, can be employed to isolate the liquid from the vacuum environment, and are useful for fundamental studies.[6-9] However, for practical applications that demand precise control of precursor concentration, diversity of substrate selection, and the creation of complex three-dimensional nanostructures, electron beam deposition in a free-surface liquid film offers many advantages and flexibility despite the challenges inherent in creating a film in vacuum. In one free-surface approach, LP-FEBID has been applied to the bulk liquid layer formed via condensation in an environmental (low vacuum) scanning electron microscope (ESEM).[10, 11] Combinations of ESEM with microwells[10] and liquid injected via nanocapillaries[12] have also been used to create thick liquid films for FEBID. In these approaches, the thickness of the liquid film has proven problematic, as the beam electrons cannot penetrate the liquid and deposits are strongly affected by the moving liquid, often failing to even adhere to the substrate. The development of Nanoelectrospray Assisted Focused Electron Beam Induced Deposition (NESA-FEBID), on the other hand, addresses this difficulty as it enables creation of a stable and ultra-thin liquid film in high vacuum.[13, 14] Recent studies using NESA FEBID have demonstrated that the choice of solvent for LP-FEBID can have significant effects on the chemistry of the process with implications for both ease and rapidity of nanostructure synthesis. For instance, the use of a pure water solvent has been shown to be challenging for creating the stable liquid films necessary for spatial control of beam-guided synthesis and can also lead to a highly oxidizing electrochemical environment, slowing growth and thus lessening the inherent advantage of LP-FEBID.

The addition of ammonia to an aqueous solution has been shown to promote nanomaterial synthesis due to its ability to make the chemical environment more reducing.[15] The presence of ammonia in water leads to an environment around the primary e-beam irradiation site that is highly reducing for metal cations (favorable for electrochemically induced metal deposition) due to (1) the switching of the behavior of the radiolytically generated hydrogen peroxide from purely oxidizing to dual reducing/oxidizing, and (2) locally depressed concentrations of short-lived oxidizing species from radiolysis consumed by $NH_3$ and $NH_2^\bullet$ that act as scavengers of oxidizing species.[15] However, high ammonia concentrations can also lead to fast growth of unwanted co-deposits away from the desired nanostructure formation location. Controlling the spatial distribution of the solid matter deposited in and around the primary electron beam irradiation site is mediated by the species transport and is crucial for achieving high resolution in electron beam synthesis for direct-write applications.

A promising strategy in spatial control of electron beam guided synthesis is to take advantage of a water-ammonia solvent that creates a reducing radiolysis environment in the immediate vicinity of primary electron beam impingement, i.e., the location of high electron dose rate (the "near-field" region). At the same time, an oxidizing environment (such as that present when pure water is the solvent) could be established outside of the primary beam impingement site or the "far-field" region. Exploiting the interplay between the redox species transport and reactions in this dual-domain environment enables an 'electrochemical lensing' effect, where the growth of nanostructures is precisely 'focused' and amplified at the near-field, while unwanted growth in the far-field is suppressed.

Uncovering and quantifying the interplay between the reaction and transport of chemical species that yields the electrochemical lensing phenomenon is the focus of this paper, which is organized as follows. First, NESA-FEBID synthesis is described, focusing on the physicochemical state of the liquid film. Then we describe relevant chemical pathways taking place in the ammonia-water solution with dissolved silver nitrate precursor during electron beam irradiation. A reaction-transport model is presented and insights from simulations are used to reveal the mechanisms resulting in the desired electrochemical lensing effect. Finally, the results of simulation are validated using corroborating experiments and generalized to broader conclusions relevant to other electrochemical synthesis methods.

**RESULTS**

**FEBID Synthesis in NESA Films**

We use a NESA-FEBID process with ammonia-water as solvent and dissolved silver nitrate ($AgNO_3$) salt as a source of cations for Ag deposition. The presence of ammonia in the solution leads to a rapid equilibrium between $Ag^+$ ions and silver diamine ion complexes, $Ag(NH_3)_2^+$, Eq (1), which favors greater $Ag(NH_3)_2^+$ concentrations with increasing ammonia content.

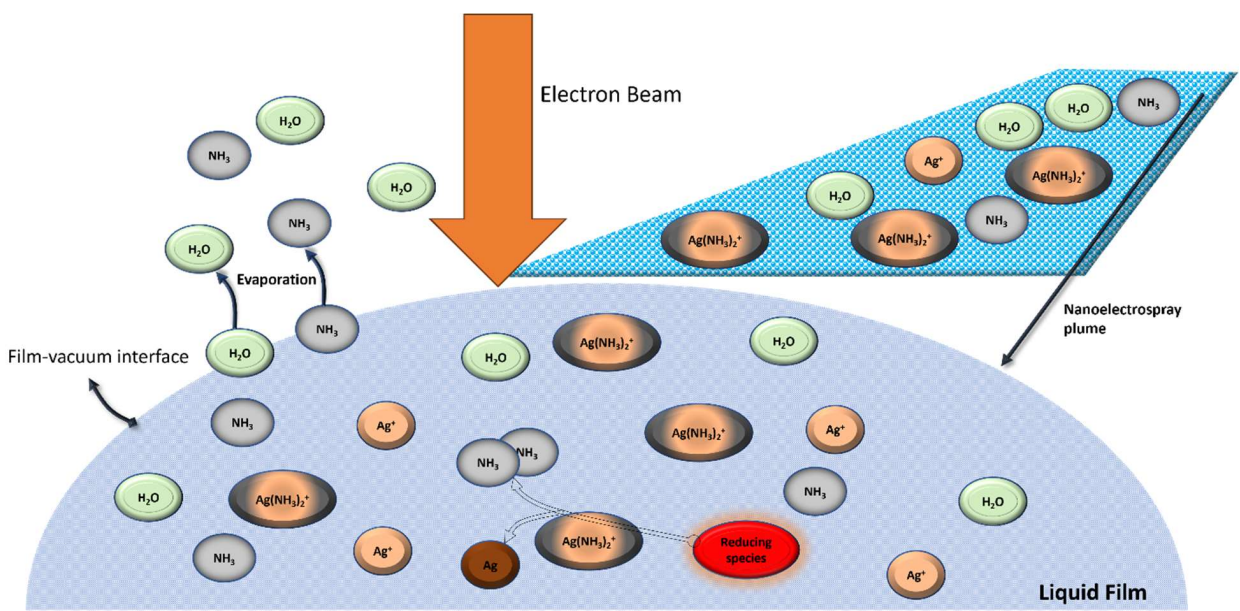

**Figure 1.** The solution of water and ammonia as solvent containing $Ag^+$ and $Ag(NH_3)_2^+$ precursor is nanoelectrosprayed onto the substrate and forms a film. As the solution is supplied to the substrate, the film size increases. Water and ammonia evaporate from the film surface slowing the growth of the film. The film reaches a quasi-steady state size and ammonia concentration as the evaporation balances the addition of solution via nanoelectrospray. The concentrations of non-volatile $Ag^+$ and $Ag(NH_3)_2^+$ precursors increase continuously (i.e., do not achieve a steady state value). The presence of $Ag(NH_3)_2^+$ provides an additional source of stored ammonia, as a single $Ag(NH_3)_2^+$ ion-complex releases two ammonia molecules if reduced to Ag.

$$Ag^+ + 2NH_{3(aq)} \rightleftharpoons Ag(NH_3)_2^+ \qquad (1)$$

In NESA-FEBID, Figure 1, irradiation of the liquid film by the electron beam results in radiolytic production of chemically reactive species. High energy electrons undergo elastic and inelastic collisions with the solvent molecules. This results in a transfer of energy from the electron beam to the liquid and leads to radiolysis, resulting in the creation of primary radiolytic species such as solvated electrons, $e_{sol}^-$, as well as $H^•$, $H_2O_2$, $HO_3^+$, $OH^•$, $HO_2^•$, $NH_2^•$ and $H_2$. These primary radiolytic species then undergo reactions with each other and the solution, leading to the creation of secondary radiolytic species such as $O_2$ and $N_2H_4$.[15, 16] Both primary and secondary radiolytic species diffuse away from their region of formation to react with the precursor, the solvent, and one another both in the near-field and far-field regions of the film.

**Reaction pathways essential to nanomaterial synthesis**

In NESA-FEBID of $AgNO_3$/ammonia-water, silver metal formation occurs through the reactions shown in Figure 2, which substantiates the impact of changing ammonia concentration. In an environment where pure water serves as the solvent (Figure 2a), an overall oxidative condition prevails, which significantly inhibits the rapid growth of nanomaterials. The species of greatest significance for synthesis of Ag nanomaterials are the ones that either reduce $Ag^+$ and $Ag(NH_3)_2^+$ to Ag or oxidize Ag to $Ag^+$. In the absence of $NH_3$, the $Ag^+$ from dissociated $AgNO_3$ is reduced to Ag by solvated electrons $e_{sol}^-$ and hydrogen radicals $H^•$. The Ag formed can then be converted back to $Ag^+$ by oxidizing species, $H_2O_2$, $OH^•$ and $O_2$. The net rate of Ag creation is determined by the difference in rates of silver cation reduction and silver atom (solid phase) oxidation (Figure 2a). The introduction of ammonia changes the deposition landscape in multiple ways (Figure 2b).

Firstly, in the presence of ammonia, the silver ions ($Ag^+$) are partially converted to silver diamine,

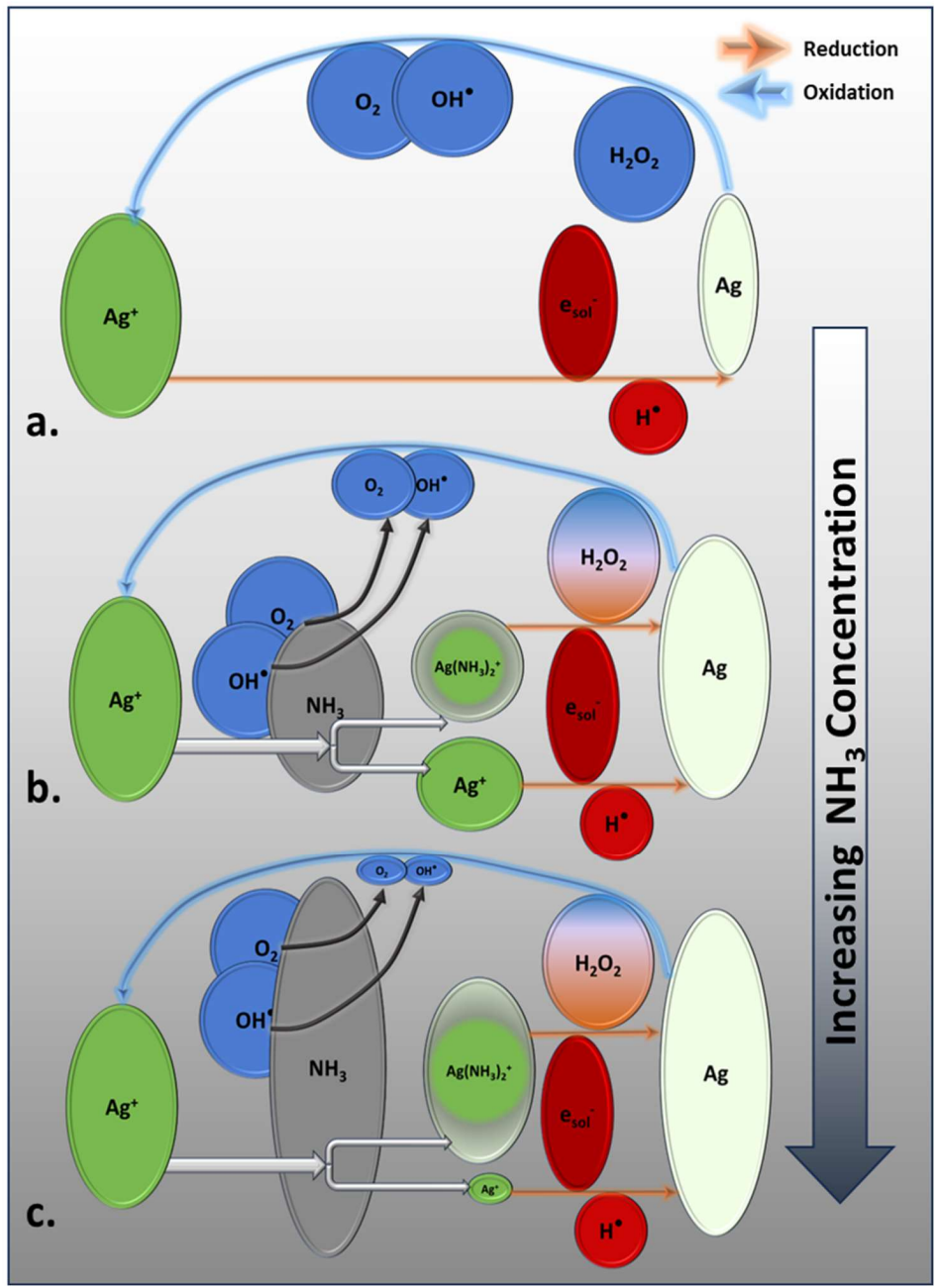

**Figure 2.** Important chemical pathways for silver production in LP-FEBID with an ammonia-water solvent and metal salt precursors (shown in green). The reducing species are illustrated in red and oxidizing species in blue. In the absence of $NH_3$ (a), the precursor $Ag^+$ is converted to Ag via reduction by solvated electrons. The Ag can then be reverted to $Ag^+$ via oxidation. The addition of ammonia to the solution (b) leads to partial conversion of $Ag^+$ to $Ag(NH_3)_2^+$. This adds an additional silver formation pathway as $Ag(NH_3)_2^+$ can be reduced to Ag by $H_2O_2$. The role of $H_2O_2$ is shifted from purely an oxidizer of Ag to both an oxidizer and reducer. Presence of $NH_3$ also depletes the concentrations of oxidizing species $O_2$ and $OH^•$, making the environment more reducing. Further addition of $NH_3$ (c) amplifies this effect, leading to the environment becoming even more reducing. The size of each species oval indicates their relative concentrations for different cases.

$Ag(NH_3)_2^+$. This introduces an additional reduction pathway to solid Ag formation.

Secondly, the role of $H_2O_2$, a primary radiolytic species, shifts from being exclusively an oxidizer for Ag to also being a reducing species. Hydrogen peroxide reduces precursor $Ag(NH_3)_2^+$ to form silver, and also reduces oxidizing species $OH^•$ and $O_2$, leading to their diminished concentration. The net effect is a significantly enhanced rate of solid Ag deposit creation, through both an increase in the reduction to Ag and a decrease in oxidation of Ag. $Ag(NH_3)_2^+$ also acts as an additional source of ammonia as each ion-complex releases two ammonia molecules into the film upon reduction (Figure 1). Increasing ammonia concentration even further amplifies these effects (Figure 2c). At high ammonia concentrations, nearly all dissolved silver precursor is in $Ag(NH_3)_2^+$, enhancing the role of $H_2O_2$ as a reducer. Additionally, the oxidizing species are highly suppressed, leading to higher net growth rates of Ag. Reaction-diffusion simulations were used to obtain detailed insights into the processes leading to variation of net reducing versus oxidating behavior as function of spatial location and time.

**Analysis of reactions and transport**

A transient mass conservation and species transport model is used to simulate the dynamic interplay between chemical reactions and the diffusion of species within the system. The liquid film, depicted at the top of the diagram in Figure 3a, is very thin owing to the nanoelectrospray delivery, such that the rate variations across it could be ignored for a quasi-1D axisymmetric treatment (Figure 3b). Rather than introduce assumptions regarding the solid phase nucleation process, we use the Ag concentration as a proxy for nanostructure creation.[15] This allows us to treat solid metal reduction and oxidation as homogeneous processes with a correction factor accounting for heterogeneous (confined to the interface) nature of oxidation. Additionally, the effects of advection and electromigration are not considered, and Fickian diffusion is assumed.

These assumptions have been shown to accurately capture the key processes and predict electron mediated metallic nanomaterial synthesis.[15]

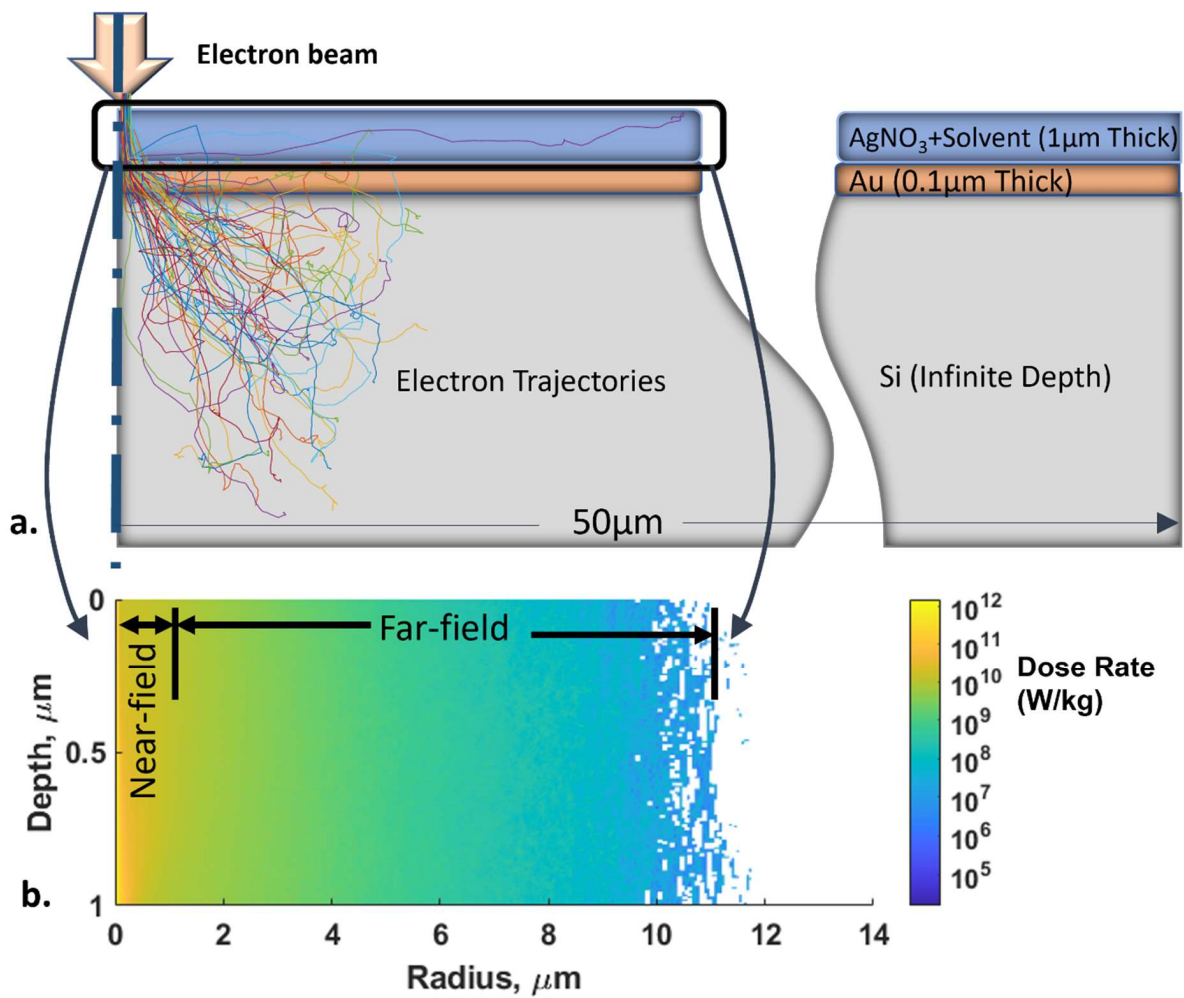

**Figure 3.** Monte-Carlo simulation is used to calculate the dose rate distribution due to a 30 kV, 3 nA electron beam irradiation. (a) The simulation domain - a 1 μm thick liquid film is overlayed on a 0.1 μm gold layer, which is placed on a thick (treated as semi-infinite) silicon substrate. (b) The axisymmetric dose rate distribution within the liquid film. The origin (0,0) is the point of electron beam irradiation, and the axis of symmetry is the vertical line passing through this point. Due to negligible axial variation in dose rate, the problem can be simplified to 1D axisymmetric.

The variation of the concentration of species $i$, $C_i$, with time and radial position, $t$ and $r$, respectively, is given by,

$$\frac{\partial C_i}{\partial t} = D_i \frac{1}{r}\frac{\partial}{\partial r}(rC_i) + S_i + R_i \tag{2}$$

$D_i$ is the diffusion coefficient of species $i$, $S_i$ is the net rate of production/consumption by homogeneous chemical reaction, and $R_i$ is the net rate of production by radiolytic processes. The homogenous reactions are treated as elementary reactions, i.e.,

$$S_i = \sum_p k_p \prod_{j \in r_p} C_j^{s_j} - \sum_c k_c \prod_{m \in r_c} C_m^{s_m} \tag{3}$$

where, $k_p$ and $k_c$ and the rate constants for the chemical reactions producing species $i$, $r_p$, and consuming species $i$, $r_c$, respectively.

The exponents $s_j$ and $s_m$ represent the associated stoichiometric coefficients. The source term representing net creation or consumption by radiolysis, $R_i$, is given by $R_i = G_i \psi$ where the G-value, $G_i$, is solvent dependent and gives the moles of species $i$ created per unit energy of electron beam absorbed, and $\psi$ is the local rate of energy deposition from e-beam irradiation, found using azimuthally and axially integrated values from Monte Carlo simulations implemented in CASINO, divided by thickness and circumference to obtain local average values for the 1-D simulations. The chemical reactions, the rate constants and G-values utilized are specified in our previous work.[15]

The reaction source term, $S_i$, as given in Eq. (3), is for homogenous reactions; however, the metallic silver produced by the NESA-FEBID process will eventually nucleate to form a nanostructure, a process that cannot be modeled explicitly in the 1-D formulations. In growing nanostructures, only the Ag present on the solid deposit surface remains exposed and susceptible for oxidation, via heterogeneous reactions. As a solid deposit increases in size its surface-to-volume ratio decreases, as does the proportion of Ag available for oxidation relative to the total amount of Ag locally present. To account for decreased availability of Ag for oxidation after nucleation, while maintaining the simplicity of a homogenous reaction model, we assume that the reaction rate of silver atoms on the surface of deposits follows the same rate expression as in the bulk, but only consider surface atoms as contributing to the concentration of Ag participating in the oxidation reaction. This approach is accomplished by modifying the form of the Ag oxidation expression with a dimensionless prefactor, $f$, that is a function of silver concentration: $f = \min(C^*/C_{Ag}, 1)$, where the value of the reference concentration $C^*$ is based on the molar density of solid silver and the film thickness (1 µm), $C^* = 38.88 \text{ mol/m}^3$.

In a NESA-FEBID, sprayed non-volatile species concentrations will continuously increase, while volatile species concentrations reach a steady state (Figure 1). We consider two initial silver salt concentrations, 26.5 mM and 265 mM, which are the concentrations occurring at 1 s and 10 s, respectively, within a film of 50 µm radius and 1 µm thickness formed via introduction of a 250 µM AgNO$_3$ solution at a rate of 3 µL/hr. Initial free (not bound in silver diamine) dissolved ammonia concentrations in the film vary from ~100 µM up to 3 M, which corresponds to the steady-state concentration achieved assuming evaporation at the kinetic limit to vacuum from a film of 50 µm radius and 1 µm thickness formed via introduction of a 5% to 30% w/w ammonia concentration in water at a rate of 3 µL/hr (typical experimental conditions). The initial concentration of silver ions and silver diamine ions is based on the total initial AgNO$_3$ concentration, the initial ammonia concentration, and the equilibrium condition, Eq. (1). Zero flux in far field and axial symmetry boundary conditions are used on the two domain boundaries.

**Electrochemical Lensing: Insights from Simulations**

Electrochemical lensing is the phenomenon in which the transport and reaction interaction result in preferential growth in the near-field while minimizing nanostructure deposition in the far-field, a phenomenon which can be exploited to achieve high resolution deposits as shown schematically in Figure 4.

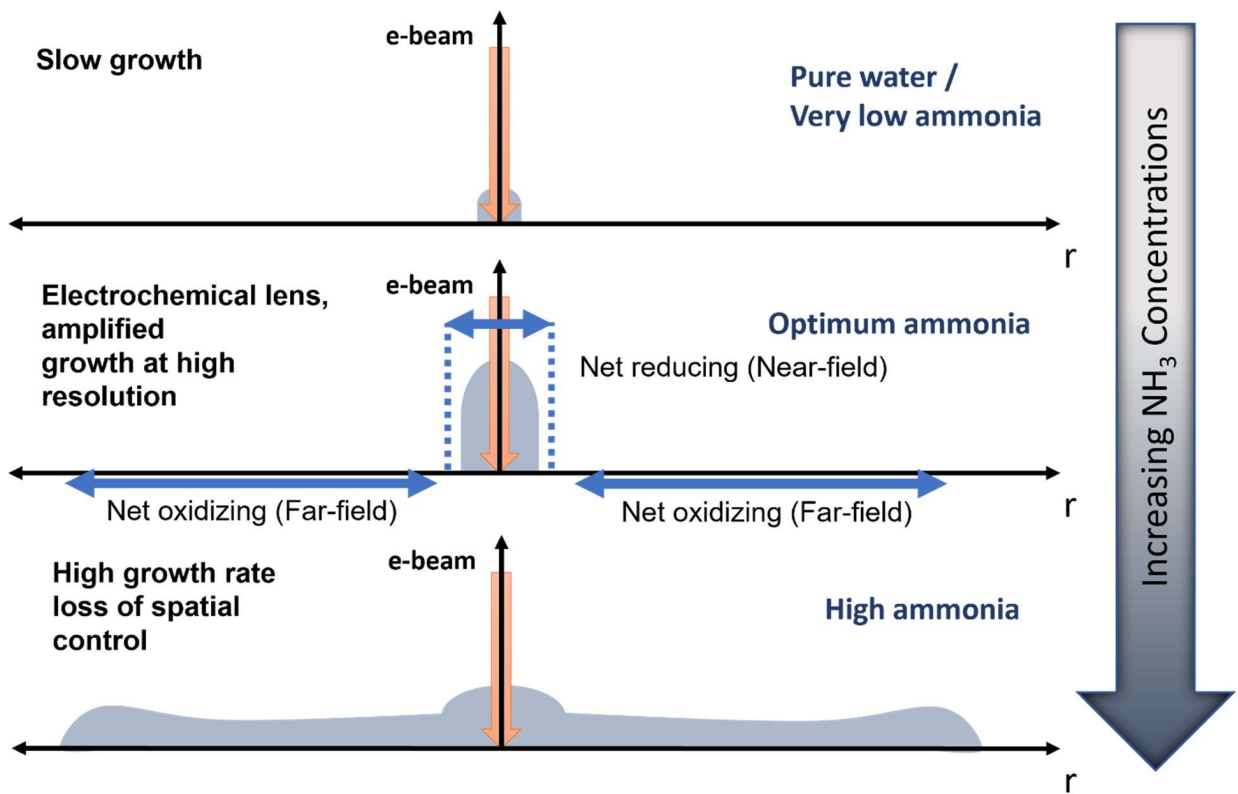

**Figure 4.** Nanostructure deposition in water-ammonia solvents. (a) In pure water or low ammonia, minimal nanostructure formation occurs due to oxidizing conditions. (b) Optimal ammonia concentrations for electrochemical lensing result in focused and rapid nanostructure growth with high resolution, as reducing conditions prevail in the near-field and oxidizing conditions prevent nanostructure formation in the far-field. (c) High ammonia concentrations lead to rapid growth of solid phase throughout the domain with loss of resolution, due to a reducing environment in both near and far-field.

To demonstrate the effect of ammonia content on deposition, we perform simulations at multiple ammonia concentrations. The key result of simulations is the concentration of Ag, which serves as a proxy for silver nanostructure growth. Figure 5 shows the resulting Ag concentrations for various NH$_3$ initial concentrations after a 1 s exposure of the film to a 30 kV electron beam.

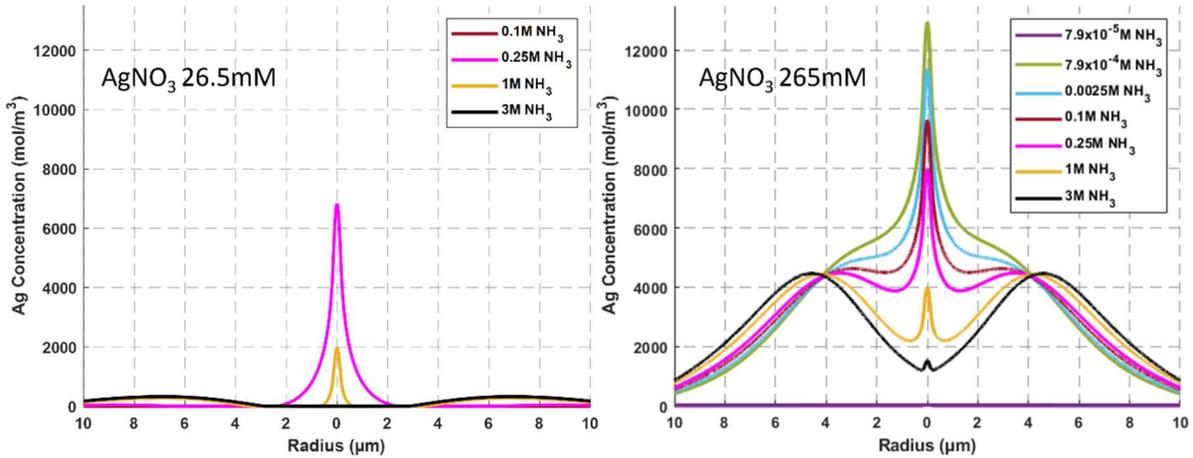

**Figure 5.** Simulated axisymmetric (mirrored around y-axis) Ag concentration profiles, after 1 second electron beam irradiation across different NH$_3$-water solution concentrations, for two initial AgNO$_3$ concentrations. Panel (a) shows the Ag concentration profile in a 26.5 mM AgNO$_3$ film, highlighting optimal NH$_3$ concentration for electrochemical lensing at 0.25 M NH$_3$; Panel (b) illustrates the 265 mM AgNO$_3$ scenario, where increased precursor availability leads to substantial far-field deposition even at very small NH$_3$ concentrations in the film, resulting in broadening growth and diminished resolution.

For the lower silver precursor concentration, 26.5 mM, electrochemical lensing is observed at an intermediate ammonia concentration (Figure 5a). Negligible silver formation is observed after 1 second of beam exposure when the initial free NH$_3$ concentration is 0.1 M: for this precursor concentration and small initial concentration of free ammonia, the entire domain remains net oxidizing for silver. When the NH$_3$ concentration is increased to 0.25 M, the near-field environment becomes reducing, and Ag growth commences at the point of electron beam impingement, while the far-field continues to be oxidizing, thus, preventing Ag production away from the e-beam irradiation. This distribution of Ag formation, resulting from creation of a reducing environment in the near-field only, is the hallmark feature of electrochemical lensing. As the NH$_3$ concentration is further increased to 1 M, both the near-field and far-field environments become reducing, resulting in broadening of the area of net silver reduction. The strongly reducing environment not only leads to a higher rate of conversion of Ag$^+$ and Ag(NH$_3$)$_2^+$ precursor ions into Ag in the far-field, but also, in turn, decreases the availability of these precursor ions for diffusion towards the near-field. Consequently, the central Ag pillar experiences a precursor shortage, slowing its growth as compared to the rate observed at 0.25 M NH$_3$. This effect is even more pronounced for higher concentrations of ammonia: for initial NH$_3$ of 3 M, significant far-field Ag formation occurs while central deposit formation is very limited.

Electrochemical lensing is not observed for all precursor silver salt concentrations. When initial AgNO$_3$ is 265 mM, even very small initial concentrations of free ammonia correspond to rapid and widespread Ag formation (Figure 5b). In part this is due to the presence of significant "bound" ammonia locked in the Ag(NH$_3$)$_2^+$ complex. When silver diamine is reduced, NH$_3$ enters solution, resulting in an increase of ammonia concentrations (Figure 1). Thus, while at extremely low initial free ammonia levels (79 µM), silver growth is negligible, high rates of near-field deposition are achieved even with only 0.79 mM of initial free ammonia. However, the lensing effect that occurs at the lower salt concentration is diminished, with significant net reduction occurring outside the near-field region, as NH$_3$ released during Ag(NH$_3$)$^+$ reduction, diffuses to the far-field. At higher ammonia concentrations (2.5 mM to 3 M), silver growth is even more pronounced in the far-field, leading to formation of volcano-like structures (Figure 5b).

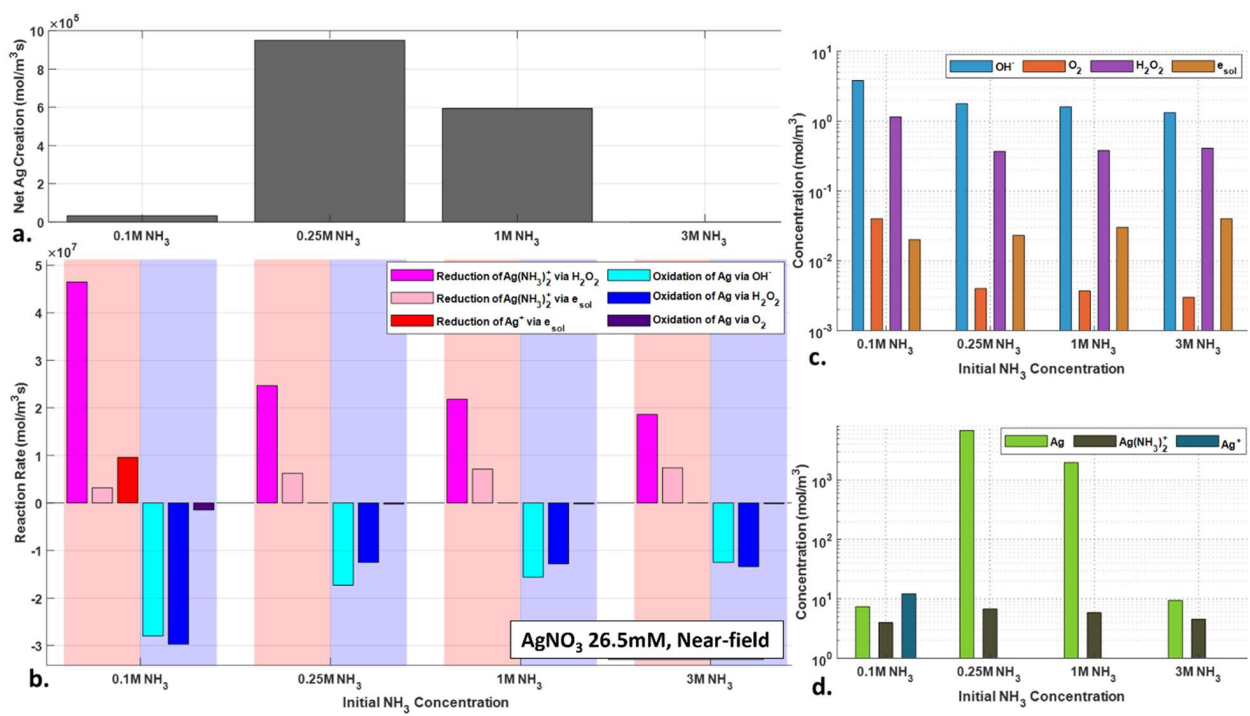

**Figure 6.** Key reaction rates and concentrations in the near-field (center) for a 26.5 mM AgNO$_3$ solution at NH$_3$ concentrations of 0.1, 0.25, 1, and 3 M post 1 sec electron beam irradiation. Panel (a) illustrates the net instantaneous rate of silver deposition showing peak Ag creation in the near-field at 0.25 M NH$_3$. The net Ag silver deposition is the sum of all the reduction and oxidation pathways for precursors and Ag, respectively (b). Concentrations of key oxidizing and reducing species are shown in (c). Concentrations of Ag and the precursor ions are shown in (d), which indicates a partial conversion of Ag$^+$ to Ag(NH$_3$)$_2$$^+$ contributing to a negligible Ag creation at 0.1 M NH$_3$ (a). Additionally, decreasing precursor concentrations from 0.25 M to 3 M NH$_3$ in (d) is responsible for the drop in Ag creation.

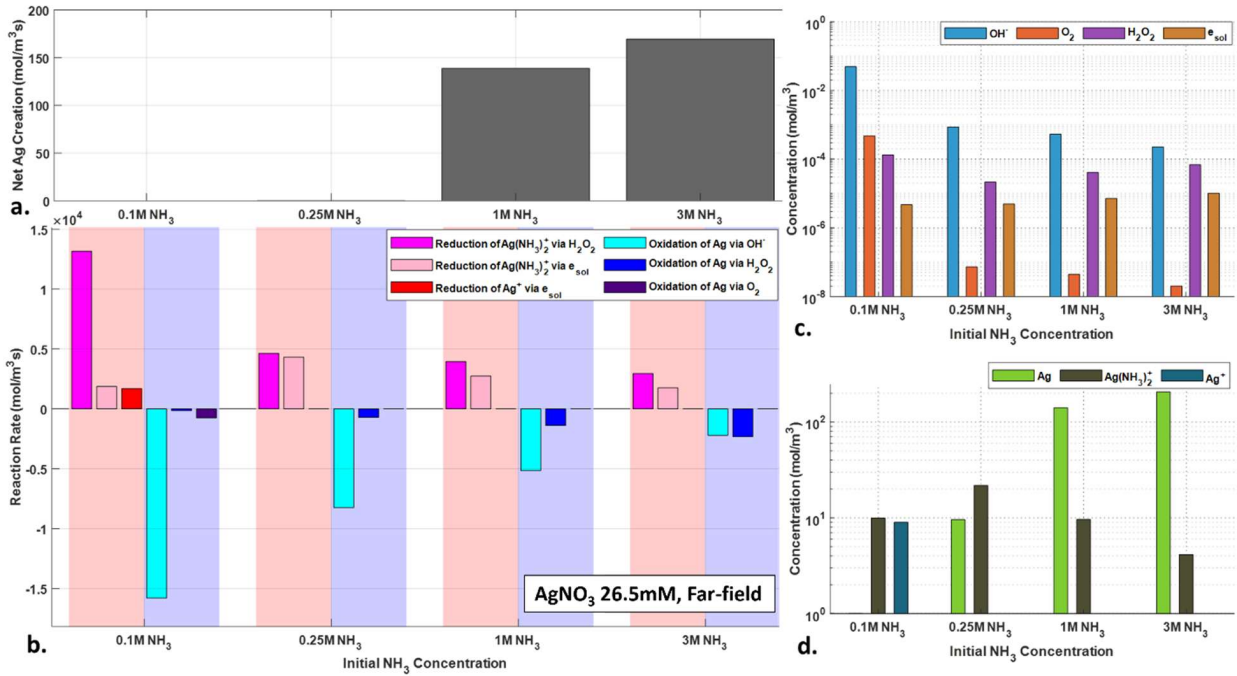

**Figure 7.** Key reaction rates and concentrations for far-field (4.5 μm from the center) silver deposition dynamics under varying $NH_3$ concentrations in a 26.5 mM $AgNO_3$ solution following electron beam exposure of 1 sec are shown. The net rate of Ag deposition (a) shows the increase in the rate of Ag creation in the far-field with higher $NH_3$ concentrations - this rate is the sum of all the reduction and oxidation reaction rates shown in (b). Concentrations of reducing and oxidizing species participating in creation or consumption of Ag is showing in (c). The Ag and precursor ions concentrations (d) show an increase in far-field Ag creation with increased $NH_3$, as well as a corresponding drop in the precursor ion concentrations.

Simulations reveal details of the fundamental principles for the observed deposition behavior as a function of ammonia and silver salt concentration. Figure 6 depicts key reaction rates and concentrations in the near-field (center) 1 sec after electron beam irradiation of a 26.5 mM silver nitrate film for four representative initial ammonia concentrations. In addition to the instantaneous rate of silver creation (Figure 6a), which mirrors the results showing net silver produced in the near-field from Figure 5a, Figure 6b depicts the dominant reduction and oxidation reaction rates.

Even at the lowest ammonia concentration, 0.1 M, solvated electron reduction of $Ag^+$ has been supplanted as the dominant Ag production route by $H_2O_2$ reduction of $Ag(NH_3)_2^+$. The higher rate of silver production at 0.25 $NH_3$ ammonia is not due to higher rates of reduction but a more significant decrease in the rate of oxidation than in the rate of reduction. In particular, the consumption of oxidizing species $OH^\bullet$ and $O_2$ by $NH_3$ and radiolytically produced $NH_2^\bullet$, as evident in Figure 6c, is responsible for the higher net silver production rate. The diminished Ag production with further increases of $NH_3$ concentrations to 1 M and 3 M corresponds with a decrease in the concentration of $Ag(NH_3)_2^+$ in the near-field, Figure 6d, which is due to consumption of the precursor during far-field silver production which prevents it from being available to diffuse to the center. Figure 7 depicts reaction rates and concentrations of key redox-mediating species in the far-field (4.5 μm from the center) for four representative initial ammonia concentrations 1 s after electron beam irradiation of a 26.5mM silver nitrate film. As was seen for moderate ammonia concentrations in the near-field, increasing rates of silver production in the far-field at higher ammonia concentrations (Figure 7a) are due primarily not to the increased rates of reduction, but as a result of even greater decrease in the rates of oxidation (Figure 7b). The oxidizing species $OH^\bullet$ and $O_2$ are formed by radiolytic processes in the near-field and must diffuse to the far-field. As $NH_3$ concentration is raised, more $OH^\bullet$ and $O_2$ are scavenged prior to reaching the far-field, Figure 7c. This leads to a faster drop in the rates of oxidation with increasing ammonia concentration than in the rates of reduction. As a result, net Ag creation occurs in the far-field for these $NH_3$ concentrations. The reducing environment in the far-field also leads to a drop in concentration of the precursor $Ag(NH_3)_2^+$ as it is consumed in the $H_2O_2$ mediated reduction reaction, Figure 7d. This leads to a decrease in the rate of $Ag(NH_3)_2^+$ diffusion into the near-field and a slower growth of Ag in the near-field.

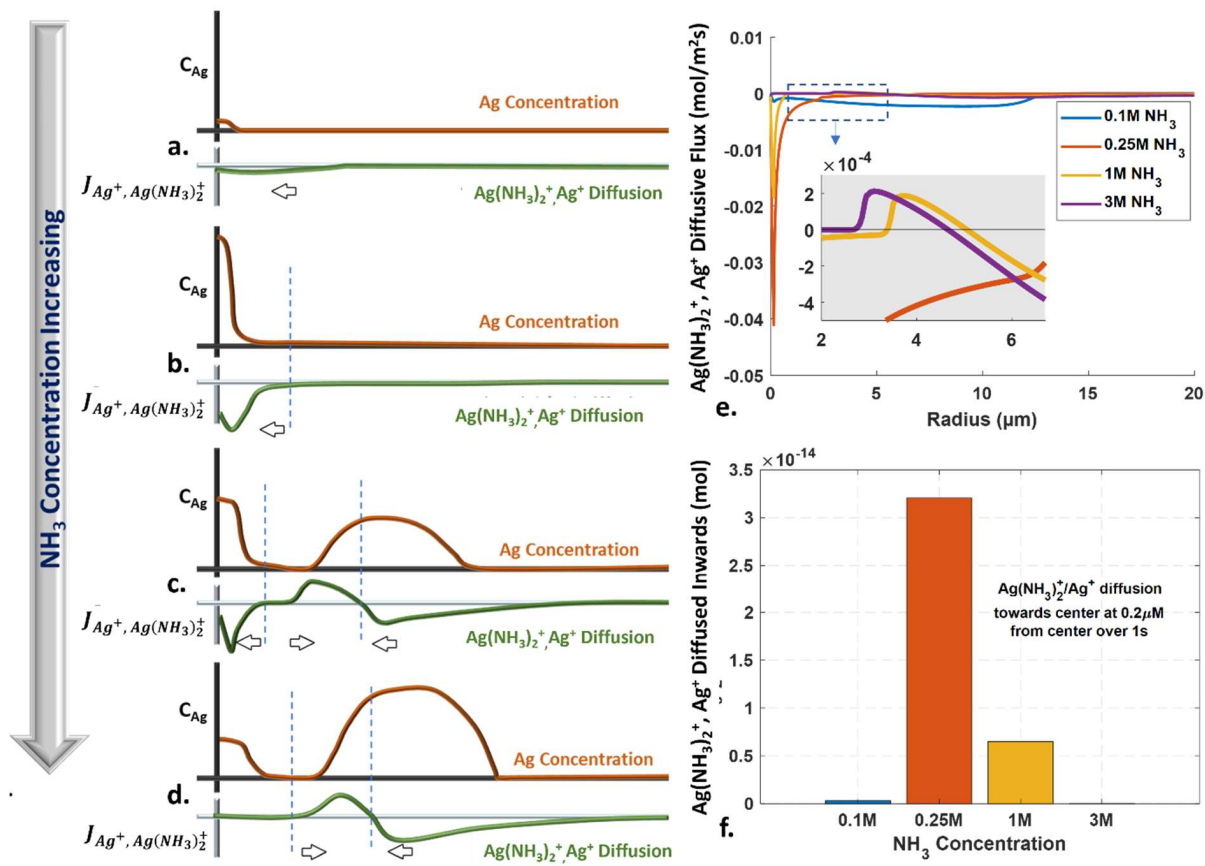

**Figure 8.** The impact of NH$_3$ concentration on Ag$^+$ and Ag(NH$_3$)$_2^+$ diffusion and Ag creation under a 30 kV electron beam for 1 second. In low NH$_3$ or pure aqueous conditions (a), oxidizing conditions inhibit Ag formation, with minimal ion diffusion. At optimal NH$_3$ concentration (b), a reducing environment in the near-field promotes precursor cations Ag$^+$ and Ag(NH$_3$)$_2^+$ conversion to Ag, while maintaining oxidizing conditions in the far-field, promoting precursor cation diffusion towards the near-field. Higher NH$_3$ concentrations (c-d) lead to Ag$^+$ and Ag(NH$_3$)$_2^+$ conversion in both the near-field and far-field, decreasing precursor cations available for diffusion towards near-field and slowing Ag growth. Panel (e) highlights the maximum diffusion of Ag$^+$ and Ag(NH$_3$)$_2^+$ towards the near-field at 0.25M NH$_3$, marking it as the optimum concentration for electrochemical lensing. Increasing NH$_3$ to 1M and 3M reduces the rates of diffusion, slowing the near-field Ag growth. Panel (f) shows the diffusive flux profiles, indicating a reversal in diffusion direction at 1 M and 3 M NH$_3$ concentrations due to the local precursor cation concentration minima.

Figure 8 depicts in greater detail the role of transport on silver deposition rates at different locations. A film consisting of extremely low (or none) $NH_3$, Figure 8a, will lead to an overall oxidizing environment. This yields very little or no creation of Ag anywhere. Additionally, very little conversion of $Ag^+$ and $Ag(NH_3)_2^+$ to Ag causes a very small concentration gradient of $Ag^+$ and $Ag(NH_3)_2^+$ and consequently a very small diffusion flux. At optimum $NH_3$ concentration, Figure 8b, reducing conditions are established at the near-field, while conditions at the far-field remain oxidizing. This leads to conversion of precursor, primarily $Ag(NH_3)_2^+$, to Ag in the near-field, with precursor supplied to the near-field by diffusion from the far-field. At higher $NH_3$ concentrations, Figure 8c and 8d, precursor is reduced to Ag both in the near-field and far-field. This leads to growth of Ag in the far-field, as well as slowing down of growth of Ag in near- field by decreasing the diffusion of precursor towards the near-field. Figure 8e shows the amount of $Ag^+$ and $Ag(NH_3)_2^+$ that diffuse towards the near-field. The ammonia concentration (0.25M) associated with the highest magnitude diffusion flux is the one at which electrochemical lensing effect occurs.

**Electrochemical Lensing - Corroborating Experiments**

To corroborate our simulation results, we conducted a series of experiments using NESA-FEBID with a 250 μM $AgNO_3$ solution, under a 30 kV electron beam for 1 second. We explored the effects for varying $NH_3$ concentrations in the solvent. Attempts to form silver deposits using $AgNO_3$ dissolved in pure water (no ammonia) were not successful, supporting the simulation predictions of suppression of Ag growth in a highly oxidizing environment. Line and square arrays of nanostructures (to ensure repeatability) deposited from initial 30%, 25% and 15% ammonia-water NESA-FEBID solutions (corresponding to 3, 1 and 0.25 M ammonia concentrations in the film) were analyzed via Atomic Force Microscopy (AFM). Figures 9a to 9i show both two- and three-dimensional images and the radial profiles of these nanostructures. To compare the experimental results with the simulations, normalized pillar heights from the experiments, Figure 9j, were

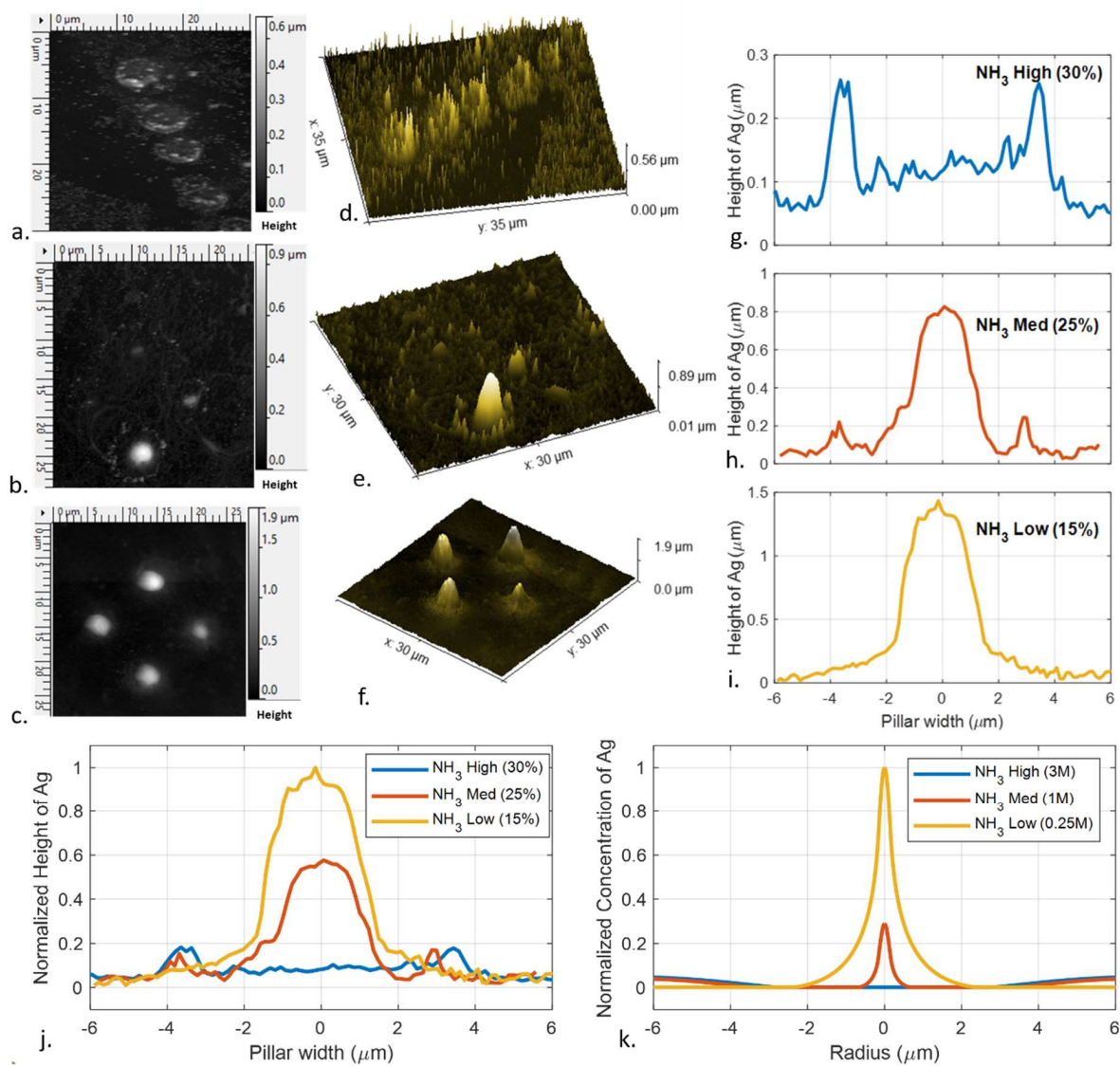

**Figure 9.** AFM images of nanostructures from NESA-FEBID using water-ammonia solutions with 250 µM AgNO$_3$ at 30%, 25%, and 15% NH$_3$ concentrations in top-view 2D (a-c) and 3D (d-f) radial profiles. Nanostructures are created with 1s 30 kV electron beam in 'spot' mode. Ag growth varies with NH$_3$ concentration, showing focused near-field growth at 15% NH$_3$ and increased far-field growth at 25% and 30% NH$_3$. Normalized experimentally determined pillar profiles (j) are compared to simulated Ag concentration profiles at varying NH$_3$ levels, showing qualitative agreement and presence of an optimal NH$_3$ concentrations for focused near-field Ag growth, the signature of the electrochemical lensing.

compared to the normalized Ag concentration profiles obtained for different film NH$_3$

concentrations from the simulations, Figure 9k.

Experimental silver creation follows the expected behavior with negligible silver growth in pure water, and a silver peak created in the near-field at 15% $NH_3$ NESA-FEBID solution with very little far-field growth. Increasing the concentration of $NH_3$ to 25% and 30% in the NESA-FEBID solution led to a higher Ag growth in the far-field and slower Ag growth in the far-field and a slower Ag growth in the near-field. This trend, predicted by simulations and confirmed by experiments, demonstrates the electrochemical lensing concept obtained when metal precursor cation is converted into its diamine complex enabling radiolytically produced $H_2O_2$ to act as a dual reducer and oxidizer. This leads to a high growth rate synthesis of high aspect ratio nanostructures in the near-field with corresponding increase in resolution. At high ammonia concentration, ammonia levels are sufficient to result in scavenging of oxidizing species, preventing them from diffusing from near-field to far-field, so that reducing species that diffuse to the far-field can react with precursor cations, producing a solid deposit, which prevents precursor from diffusing into the near-field. At lower ammonia levels, lesser conversion of precursor cation to its diamine form, keeps $H_2O_2$ acting mainly as an oxidizer. This results in higher oxidation rates than reduction, limiting nanostructure formation.

**Conclusion**

We present an approach to resolving the fundamental tradeoff between the growth rate and resolution in electron beam mediated direct-write nanofabrication through the technique of electrochemical lensing. When deposition processes are governed by systems with a linear behavior, the increase in concentration of a reactant, even locally at the desired deposition location, will result in a broader deposition region due to diffusion. Electrochemical lensing is accomplished via introduction of intentional non-linearity into the system. Metal nanostructure synthesis using

FEBID from silver salt in a water-ammonia solution is a demonstration of electrochemical lensing, in which the strong non-linearity of the system's response to changing ammonia concentration comes from the multiple roles played by ammonia in the system: (1) by binding silver cations into diamine silver complexes, ammonia adds a new reduction pathway involving hydrogen peroxide and provides a mechanism for ammonia storage in the solution; (2) by using both ammonia and its radiolytic decomposition product $NH_2^\bullet$ to selectively scavenge oxidizing species. Generalization of the approach therefore requires seeking similarly complementary nonlinearities for exploitation either through use of a solvent for forming multiple forms of a cation whose reduction leads to solid formation and/or encoding the redox switchable behavior to one or more transient intermediary species that facilitates the synthesis reaction. Electrochemical lensing as a conceptual model for synthesis improvement could play a significant role in a wide variety of nanomanufacturing applications, offering a scalable, precise, and efficient pathway for the creation of nanostructures with greater control of shape and resolution.

**Methods**

HPLC grade deionized water (Sigma Aldrich) and an HPLC grade water-ammonia solution (Fisher) were used for all experiments. Silver nitrate salt (Sigma Aldrich) was dissolved in water and mixed with water-ammonia to achieve desired concentrations. A TA Hamilton 1750 TLL syringe was used to deliver the solution at 3 μL/hr via a syringe pump. The solution nanoelectrospray was carried out using a fused silica capillary emitter (360 μm outer diameter, 100 μm inner diameter, pulled to a 3±1 μm diameter tip using a Sutter P-2000 Laser-Based Micropipette Puller). The substrate was a 1 cm x 1 cm square Si wafer coupon with a 100 nm gold coating. The capillary to substrate distance is maintained at ~200 μm. The negative mode nanoelectrospray was initiated by electrically biasing the solution with -500 V at the metallic union using a SRS 5 kV power supply. Experiments were conducted using the FEI Quanta 200 SEM operated in high vacuum mode (~$10^{-4}$ torr). To distinguish nanostructure deposits from salt precipitates, Nanometer Pattern Generation System (NPGS) exposed the film in a predefined pattern of four 1-second spots, facilitating the formation of discernible nanostructures. Post-deposition, the samples were immersed in water for 2 hours and subsequently dried. The Bruker Dimension Icon was used to perform AFM imaging. Monte Carlo simulations were executed using the CASINO software. The model involving the reaction and transport of species was implemented in COMSOL Multiphysics software.

AUTHOR INFORMATION

## Corresponding Author

**Andrei G. Fedorov,** George W. Woodruff School of Mechanical Engineering, Georgia Institute of Technology, 771 Ferst Dr NW, Atlanta, GA 30332, USA, https://orcid.org/0000-0003-0859-2541, email: AGF@gatech.edu.

**Authors**

**Auwais Ahmed,** George W. Woodruff School of Mechanical Engineering, Georgia Institute of Technology, 771 Ferst Dr NW, Atlanta, GA 30332, USA, https://orcid.org/0000-0002-1120-985X

**Peter A. Kottke,** George W. Woodruff School of Mechanical Engineering, Georgia Institute of Technology, 771 Ferst Dr NW, Atlanta, GA 30332, USA, https://orcid.org/0000-0003-1382-5215

**Author Contributions**

Conceptualization: A.A., P.A.K., and A.G.F. Methodology: A.A., P.A.K., and A.G.F. Software: A.A., Visualization: A.A., Investigation: A.A., P.A.K and A.G.F., Resources: A.G.F., Writing A.A., P.A.K., and A.G.F., Supervision: A.G.F. Project administration: A.G.F. Funding acquisition: A.G.F.

ACKNOWLEDGMENT

This work was supported by the U.S. Department of Energy (DOE), Office of Science, Basic Energy Sciences (BES), under Award #DE-SC0010729.